\documentclass{article}
\usepackage{epsfig}
\usepackage[hang,nooneline]{subfigure}
\date{\today}

\usepackage{amsmath}
\usepackage{amssymb}

\begin{document}
\title{A scalar field instability of rotating and charged black holes
in (4+1)-dimensional Anti-de Sitter space-time}
\author{{\large Yves Brihaye \footnote{email: yves.brihaye@umons.ac.be} }$^{\ddagger}$ and 
{\large Betti Hartmann \footnote{email: b.hartmann@jacobs-university.de}}$^{\dagger}$
\\ \\
$^{\ddagger}${\small Physique-Math\'ematique, Universite de
Mons-Hainaut, 7000 Mons, Belgium}\\ 
$^{\dagger}${\small School of Engineering and Science, Jacobs University Bremen, 28759 Bremen, Germany}  }

\date{}
\newcommand{\dd}{\mbox{d}}
\newcommand{\tr}{\mbox{tr}}
\newcommand{\la}{\lambda}
\newcommand{\ka}{\kappa}
\newcommand{\f}{\phi}
\newcommand{\vf}{\varphi}
\newcommand{\F}{\Phi}
\newcommand{\al}{\alpha}
\newcommand{\ga}{\gamma}
\newcommand{\de}{\delta}
\newcommand{\si}{\sigma}
\newcommand{\bomega}{\mbox{\boldmath $\omega$}}
\newcommand{\bsi}{\mbox{\boldmath $\sigma$}}
\newcommand{\bchi}{\mbox{\boldmath $\chi$}}
\newcommand{\bal}{\mbox{\boldmath $\alpha$}}
\newcommand{\bpsi}{\mbox{\boldmath $\psi$}}
\newcommand{\brho}{\mbox{\boldmath $\varrho$}}
\newcommand{\beps}{\mbox{\boldmath $\varepsilon$}}
\newcommand{\bxi}{\mbox{\boldmath $\xi$}}
\newcommand{\bbeta}{\mbox{\boldmath $\beta$}}
\newcommand{\ee}{\end{equation}}
\newcommand{\eea}{\end{eqnarray}}
\newcommand{\be}{\begin{equation}}
\newcommand{\bea}{\begin{eqnarray}}

\newcommand{\ii}{\mbox{i}}
\newcommand{\e}{\mbox{e}}
\newcommand{\pa}{\partial}
\newcommand{\Om}{\Omega}
\newcommand{\vep}{\varepsilon}
\newcommand{\bfph}{{\bf \phi}}
\newcommand{\lm}{\lambda}
\def\theequation{\arabic{equation}}
\renewcommand{\thefootnote}{\fnsymbol{footnote}}
\newcommand{\re}[1]{(\ref{#1})}
\newcommand{\R}{{\rm I \hspace{-0.52ex} R}}
\newcommand{\N}{{\sf N\hspace*{-1.0ex}\rule{0.15ex}%
{1.3ex}\hspace*{1.0ex}}}
\newcommand{\Q}{{\sf Q\hspace*{-1.1ex}\rule{0.15ex}%
{1.5ex}\hspace*{1.1ex}}}
\newcommand{\C}{{\sf C\hspace*{-0.9ex}\rule{0.15ex}%
{1.3ex}\hspace*{0.9ex}}}
\newcommand{\eins}{1\hspace{-0.56ex}{\rm I}}
\renewcommand{\thefootnote}{\arabic{footnote}}

\maketitle

\bigskip

\begin{abstract}
We study the stability of static as well as of rotating and charged black holes in (4+1)-dimensional
Anti-de Sitter space-time which possess spherical horizon topology. 
We observe an instability related to the condensation of a charged, tachyonic scalar field
and construct ``hairy'' black hole solutions of the full non-linear system of coupled
Einstein, Maxwell and scalar field equations. We observe that the limiting
solution for small horizon radius is either
a hairy soliton solution or a singular solution that is not a regular extremal solution.
 Within the context of the gauge/gravity duality the condensation of the scalar
field describes a holographic conductor/superconductor phase transition on the surface of a sphere.
\end{abstract}
\section{Introduction}

In recent years the gravity/gauge theory duality \cite{ggdual} and in particular
the AdS/CFT correspondence \cite{adscft} has been used to understand phenomena such as
superconductivity and superfluidity with the help of black holes in higher dimensions 
\cite{gubser,hhh,reviews}.
This approach relies on the fact that the AdS/CFT correspondence states
that a gravity theory in a $d$-dimensional
Anti--de Sitter (AdS) space--time is equivalent to a Conformal Field Theory (CFT) 
on the $(d-1)$-dimensional boundary of AdS. This duality is a weak-strong coupling duality
which allows to study strongly coupled Quantum Field Theories with the help
of black hole and soliton solutions of standard General Relativity. 
The conductor/superconductor
phase transitions are described by the onset of an instability of a black hole
solution to a gravity-U(1) gauge theory
with respect to the condensation of a tachyonic charged scalar field. 
Close to the horizon of the black hole the effective mass of the scalar field can become
negative with masses below the Breitenlohner--Freedman bound \cite{bf} such that the scalar
field becomes unstable and possesses a non--vanishing value on and close to the horizon
of the black hole. The local bulk U(1) symmetry  is associated to a global U(1) symmetry on the AdS boundary and
the value of the scalar field on the AdS boundary  with the corresponding condensate in the dual theory.
The Hawking temperature of the black hole is identified with the temperature of the dual theory. 
For temperatures above a critical value $T_c$ the black hole possesses no scalar hair and is the gravity dual of a conductor, while for 
temperatures below $T_c$ 
the black hole has scalar hair and corresponds to a superconductor.

These models can be extended to describe holographic fluid/superfluid
phase transitions by making the spatial component of the U(1) gauge field in the bulk theory 
non-vanishing which
corresponds to non-vanishing superfluid currents on the AdS boundary 
\cite{hks,superfluids,hartmann_brihaye1,hartmann_brihaye2,sonnerwithers}.

In most studies the AdS black holes are static and the dual theory hence describes non--rotating
superconductors. In \cite{sonner} a Kerr--Newman--AdS black hole \cite{carter1} has been considered in order to 
describe (2+1)--dimensional 
rotating superconductors living on the surface of a sphere and it was shown 
that a critical value of
the rotation parameter exists such that superconductivity gets destroyed. 
This was interpreted as being analogous
to a critical magnetic field.

In order to describe insulator/superconductor phase transitions soliton solutions in higher dimensional
AdS have been used \cite{nrt, horowitz_way}. 
The AdS soliton is a double Wick rotated black hole, where one of the 
coordinates needs to be compactified to a circle in order to avoid a conical singularity.
The AdS soliton has been used before in the context of the description of a confining vacuum 
in the dual gauge theory \cite{witten2,horowitz_myers} since it possesses a mass gap. 
While for spherically symmetric
black holes in AdS there is the Hawking-Page phase transition from the AdS black hole to global AdS space-time when
lowering the temperature \cite{hawking_page}, this is different for AdS black holes 
with Ricci-flat horizons used in the description of holographic superconductors.
In that case, there is a phase transition between the AdS black hole and the AdS soliton \cite{ssw} which was interpreted as
a confining/deconfining phase transition in the dual gauge theory. 
Now taking the viewpoint of  condensed
matter this phase transition describes a 1st order insulator/conductor phase transition. 
To complete the picture it was realized in \cite{nrt} that one can add a chemical potential to the AdS soliton.
Increasing the chemical potential $\mu$ the AdS soliton becomes unstable to the formation of scalar hair above some
critical value $\mu_{cr}$. This was interpreted as a 2nd order insulator/superconductor phase transition that is possible
even at zero temperature. 

While in the description of holographic superconductors or superfluids the fact that the
scalar field is charged under the U(1) is vital for the onset of the instability, 
condensation of uncharged
scalar fields in the $d$-dimensional planar Reissner-Nordstr\"om-AdS (RNAdS)
black hole 
space-time \cite{hhh} has also been observed. 
This is related to a new type of instability that is not connected
to a spontaneous symmetry breaking as in the charged case.
Rather it is related to the fact that the planar RNAdS black hole possesses an
extremal limit
with vanishing Hawking temperature and near-horizon geometry 
AdS$_2\times {\mathbb{R}}^{d-2}$ (with $d \ge 4$)
\cite{Robinson:1959ev,Bertotti:1959pf,Bardeen:1999px}. For scalar field masses 
larger than the $d$-dimensional BF bound, but smaller than the $2$-dimensional
BF bound the near-horizon geometry
becomes unstable to the formation of scalar hair, while the asymptotic AdS$_d$
remains stable \cite{hhh}.
The fact that the near-horizon geometry of extremal black holes is a topological
product of two manifolds with constant
curvature has let to the development of the entropy function formalism
\cite{Sen:2005wa,sen2,dias_silva}. 
In \cite{dias} the question of the condensation of an uncharged scalar
field on uncharged black holes in $(4+1)$ dimensions has been addressed.
In particular rotating uncharged black holes were discussed. 
It was shown that only very cold black holes are unstable to the condensation
of a tachyonic scalar field. The instability of static and uncharged black holes with
hyperbolic horizon topology was also discussed \cite{dias}. This was extended
to include Gauss-Bonnet corrections as well as solutions that possess a finite
number of nodes of the scalar field function \cite{hartmann_brihaye3}.   

While the present paper was finalized a preprint \cite{dias2} appeared in which
static, spherically symmetric black hole and soliton solutions to Einstein-Maxwell 
theory coupled to a charged, massless scalar field 
in (4+1)-dimensional AdS space-time have been studied. The existence
of solitons in global AdS was discovered in \cite{basu}, 
where a perturbative approach was taken. In \cite{dias2} it was shown that
solitons can have arbitrarily large charge for large enough gauge coupling, 
while for small gauge coupling the solutions exhibit a spiralling behaviour towards
a critical solution with finite charge and mass. The stability of RNAdS
solutions was also studied in this paper. It was found that for small gauge coupling RNAdS black holes
are never unstable to condensation of a massless, charged scalar field, while for 
intermediate gauge couplings RNAdS black hole become unstable for sufficiently large charge.
For large gauge coupling RNAdS black holes are unstable to formation of massless scalar hair for
all values of the charge. Moreover, it was observed that for large gauge coupling and
small charges the solutions exist all the way done to vanishing horizon. The limiting
solutions are the soliton solutions mentioned above. On the other hand for large charge the
limiting solution is a singular solution with vanishing temperature and finite entropy, which
is not a regular extremal black hole \cite{fiol1}.

In this paper, we are interested in the condensation of a charged tachyonic scalar 
on charged black holes in (4+1)-dimensional Anti-de Sitter space-time. We study both non-rotating, i.e.
static solutions as well as rotating solutions with spherical horizon topology. In the case of non-rotating solutions
explicit solutions without scalar fields are known - the Reissner-Nordstr\"om-Anti-de Sitter (RNAdS)
solutions. However, in the case of rotating solutions even without scalar
hair only uncharged solutions are known in closed form. These are the AdS generalizations
\cite{HHT,GLP} of the Myers-Perry black holes
\cite{MP}. While gauged supergravity models possess rotating and charged
black hole solutions \cite{sugra_bh}, rotating black hole
solutions of pure Einstein-Maxwell theory are not known in closed form. These
have been constructed numerically in the asymptotically flat case \cite{kunz1}
as well as in AdS for equal angular momenta \cite{kunz2} and with higher order
curvature corrections in the uncharged case \cite{brihaye_radu} and in the charged 
case \cite{brihaye1}, respectively. 
We will hence have to construct the rotating and charged black holes with and without scalar
hair numerically. 

Our paper is organized as follows: we present the model in Section 2. We then discuss static
black holes in Section 3 and rotating black holes in Section 4. We conclude and summarize in Section 5.

\section{The model}
In this paper, we are studying the formation of scalar hair on a
charged black hole in $(4+1)$-dimensional Anti--de Sitter space--time. 
The action reads~:
\begin{equation}
\label{action}
S= \frac{1}{\gamma} \int d^5 x \sqrt{-{\rm det}(g_{\mu\nu})} \left(R -2\Lambda 
+ \gamma {\cal L}_{\rm matter}\right) \ ,
\end{equation}
where $\Lambda=-6/L^2$ is the cosmological constant and $\gamma=16\pi G$ with $G$ Newton's constant.
${\cal L}_{\rm matter}$ denotes the matter Lagrangian~:
\begin{equation}
{\cal L}_{\rm matter}= -\frac{1}{4} F_{MN} F^{MN} - 
\left(D_M\psi\right)^* D^M \psi - m^2 \psi^*\psi  \ \ , \ \  M,N=0,1,2,3,4  \ ,
\end{equation}
where $F_{MN} =\partial_M A_N - \partial_N A_M$ is the field strength tensor and
$D_M\psi=\partial_M \psi - ie A_M \psi$ is the covariant derivative.
$e$ and $m$ denote the electric charge and mass of the scalar field $\psi$, respectively.

In the following, we want to study rotating black holes in $4+1$ dimensions. In general, such a solution
would possess two independent angular momenta associated to the two independent planes of 
rotation.
Here, we will restrict ourselves to the case where these two angular momenta are equal
to each other. The event horizon then possesses spherical topology and 
the Ansatz for the metric reads \cite{kunz1}
\begin{eqnarray}
ds^2 & = & -b(r) dt^2 + \frac{1}{f(r)} dr^2 + g(r) d\theta^2 + h(r)\sin^2\theta \left(d\varphi_1 - 
\omega(r) dt\right)^2 + h(r) \cos^2\theta\left(d\varphi_2 -\omega(r)dt\right)^2 \nonumber \\
&+& 
\left(g(r)-h(r)\right) \sin^2\theta \cos^2\theta (d\varphi_1 - d\varphi_2)^2 \ ,
\end{eqnarray}
where $\theta$ runs from $0$ to $\pi/2$, while $\varphi_1$ and $\varphi_2$ are 
in the range $[0,2\pi]$.
The solution possesses two rotation planes at $\theta=0$ and $\theta=\pi/2$ and the isometry
group is $\mathbb{R}\times U(2)$.
In the following, we will additionally fix the residual gauge freedom by choosing $g(r)=r^2$. 

For the electromagnetic field and the scalar field we choose~:
\begin{equation}
A_{M}dx^M = \phi(r) dt  + A(r)\left(\sin^2\theta d\varphi_1 + \cos^2\theta d\varphi_2\right) \  \  
\  , \   \   \   \psi=\psi(r)  \ .
\end{equation}
Note that originally the scalar field is complex, but that we can gauge away the non-trivial
phase and choose the scalar field to be real.
The coupled Einstein and Euler--Lagrange equations are obtained from the variation of the
action with respect to the matter and metric fields, respectively. These
depend on four independent constants: Newton's constant $G$, 
the cosmological constant $\Lambda$ (or Anti-de Sitter radius $L$) and the charge $e$ and mass $m$ of the 
scalar field. Here, we set $m^2 = - 3/L^2$. The system possesses two scaling symmetries:
\begin{equation}
 r\rightarrow \lambda r \ \ \ , \ \ \ t\rightarrow \lambda t \ \ \ , \ \ \ L\rightarrow \lambda L
\ \ \ , \ \ \ e\rightarrow e/\lambda \ \ \ , \ \ \ A(r) \rightarrow \lambda A(r) \ \ \ ,  \ \ \ 
h(r) \rightarrow \lambda^2 h(r)  
\end{equation}
as well as 
\begin{equation}
 \phi \rightarrow \lambda \phi \ \ \ , \ \ \ \psi \rightarrow \lambda \psi \ \ \ , \ \ \ 
A(r) \rightarrow \lambda A(r) \ \ \ , \ \ \ 
e\rightarrow e/\lambda \ \ \ ,  \ \ \ \gamma\rightarrow \gamma/\lambda^2
\end{equation}
which we can use to set $L=1$ and $\gamma$ to some fixed value without loosing generality.

Note that the metric on the AdS boundary in our model is that of a static 4-dimensional
Einstein universe with boundary metric $\gamma_{\mu\nu}$, $\mu$, $\nu=0,1,2,3$ given by
\begin{equation}
 \gamma_{\mu\nu}dx^{\mu} dx^{\nu} = -dt^2 + L^2 \left(d\theta^2 + \sin^2\theta d\varphi_1^2 + \cos^2\theta d\varphi_2^2\right)
\end{equation}
and is hence non-rotating. This is different to the case of rotating and charged black holes
in 4-dimensional AdS space-time, the so-called Kerr-Newman-AdS solutions \cite{carter1}, which
possess a boundary theory with non-vanishing angular velocity. These have been used to
describe rotating holographic superconductors \cite{sonner} and it was shown that the larger
the angular momentum of the black hole the lower is the critical temperature at which condensation
sets in. This was related to the fact that a rotating superconductor generates a magnetic field,
the so-called London field and the observation that superconductivity ceases to exist above
a critical magnetic field.

\subsection{Boundary conditions}
We are interested in solutions possessing a regular horizon at $r=r_h$.
Hence we require
\be
\label{boundary_cond} 
f(r_h)=0 \ \ , \ \  b(r_h) = 0 \ \ , \ \ \ee
such that the metric fields have the following behaviour close to the horizon \cite{brihaye1}
\begin{eqnarray}
 f(r)&=&f_1(r-r_h)+O(r-r_h)^2 \ \ \ , \ \ \ b(r)=b_1(r-r_h)+O(r-r_h)^2 \ \ \ , \nonumber \\
h(r)&=&h_0 + O(r-r_h) \ \ \ , \ \ \ \omega(r)=\omega(r_h) + w_1(r-r_h) + O(r-r_h)^2 \ \ , 
\end{eqnarray}
where $\omega(r_h)\equiv \Omega$ corresponds to the angular velocity at the horizon
and $f_1$, $b_1$ and $h_0$ are constants that have to be determined numerically. 
In addition there is a regularity condition for the metric fields on the horizon
given by $\Gamma_1(f,b,b',h,h',\omega,\omega')=0$, where $\Gamma_1$ is a lengthy polynomial
which we do not give here.
For the matter fields we find that 
\begin{equation}
 \left. \left(\phi(r) + A(r) \omega(r)\right)\right\vert_{r=r_h}=0 \ .
\end{equation}
On the AdS boundary the fields have the following behaviour \cite{brihaye_radu}
\begin{eqnarray}\
\label{bcads1}
 f(r >>1)&=&1+\frac{r^2}{L^2} + \frac{f_2}{r^2} +  O(r^{-4}) \ \ , \ 
\ b(r>>1)=1+\frac{r^2}{L^2} + \frac{b_2}{r^2} + O(r^{-4}) \ \ , \nonumber \\
h(r>>1)&=&r^2 + L^2 \frac{f_2-h_2}{r^2} + O(r^{-6}) \ \ , \ \  \omega(r>>1)=\frac{\omega_4}{r^4} + O(r^{-8}) \ ,
\end{eqnarray}
for the metric fields and
\begin{eqnarray}
\label{bcads2}
\phi(r>>1)=\mu + \frac{Q}{r^2} + O(r^{-4}) \ \ ,  \ \  
A(r>>1)=\frac{Q_m}{r^2} + O(r^{-4}) \ \ ,  \ \  
  \psi(r  >>1 ) =  \frac{\psi_{-}}{r} + \frac{\psi_{+}}{r^{3}} 
\end{eqnarray}
for the matter fields, where $Q$ and $Q_m$ are related to the electric and magnetic charge 
of the solution, respectively. $\mu$ is a constant that within the context of the gauge/gravity
duality can be interpreted as chemical potential of the boundary theory.  
For the rest of
the paper we will choose $\psi_{-}=0$ and interpret $\psi_+$ as the value of the ``condensate'', i.e.
the expectation value of the boundary operator
in the dual theory. In fact, in applications of the AdS/CFT correspondence 
appropriate powers of $\psi_+$ are typically used.

\subsection{Physical properties}
The temperature of the black hole is given by
$T=\frac{\kappa}{2\pi}$, 
where $\kappa$ is the surface gravity with
\begin{equation}
 \kappa^2=\left. -\frac{1}{2} (D_{\mu} \chi_{\nu}) (D^{\mu} \chi^{\nu})\right\vert_{r_h} \ .
\end{equation}
and $\chi=\partial_t + \Omega (\partial_{\varphi_1}+\partial_{\varphi_2})$ is the Killing vector
that is orthogonal to and null on the horizon. Moreover, the entropy is given by
$S=A/(4G)$, where $A$ is the horizon area. Using the expansions of the metric functions, we find that \cite{brihaye_radu}
\begin{equation}
 T=\frac{\sqrt{f_1 b_1}}{4\pi} \ \ , \ \ S=\frac{V_3}{4G} r_h^2 \sqrt{h_0} \ , 
\end{equation}
where $V_3=2\pi^2$ denotes the area of the three-dimensional sphere with unit radius.
The energy $E$ and angular momentum $\bar{J}$ are \cite{brihaye_radu}
\begin{equation}
 E=\frac{V_3}{8\pi G} 3 M \ \ {\rm with} \ \ M=\frac{f_2 - 4 b_2}{6} \ \ , \ \ 
\bar{J}=\frac{V_3}{8\pi G} J \ \ {\rm with} \ \ J=\omega_4   \ .
\end{equation}

\section{Static black holes}
This corresponds to the limit $\omega(r)\equiv 0$, $A(r)\equiv 0$ and $g(r)=h(r)=r^2$. 
Hence, the black holes carry only electric charge. In the following, we will reparametrize the
metric and choose $b(r)=f(r) a^2(r)$. The coupled Euler-Lagrange and Einstein equations then
read
\begin{eqnarray}
\label{eq1}
     f' &=& \frac{2}{r} \left(1-f+ \frac{2 r^2}{L^2}\right)
     - \frac{\gamma}{3}r
     \left(\frac{e^2 \phi^2 \psi^2}{f a^2} +  m^2 \psi^2 + \frac{\phi'^2}{2 a^2} + 
f \psi'^2\right)  \ , \\
\label{eq2}
        a' &=& \frac{\gamma}{3} r \left(\frac{e^2\phi^2 \psi^2}{a f^2} + a  \psi'^2\right) \ , \\
\label{eq3}
   \phi'' &=& - \left(\frac{3}{r} - \frac{a'}{a}\right) \phi' +2 \frac{e^2 \psi^2}{f} \phi \ , \\
\label{eq4}
    \psi'' &=& -\left(\frac{3}{r} + \frac{f'}{f} + \frac{a'}{a}\right) \psi'- 
\left(\frac{e^2 \phi^2}{f^2 a^2} - \frac{m^2}{f}\right) \psi \ ,
\end{eqnarray}
where now and in the following the prime denotes the derivative with respect to $r$.
Note that we will choose $m^2=-3/L^2$ which is well above the Breitenlohner-Freedman (BF) bound 
\cite{bf} $m^2_{\rm BF}=-4/L^2$. However, as has been noted first by Gubser \cite{gubser}
the coupling to the electromagnetic field renders an effective mass of the scalar field
given by $m_{\rm eff}^2=m^2-e^2\phi^2/(f a^2)$. Close to the horizon $1/f$ can become
very large and thus the effective mass can drop below the BF bound producing an instability
and henceforth scalar condensation close to the black hole horizon.

\subsection{Known solutions}
In the case where the scalar field is zero, $\psi(r)\equiv 0$, the solution is a
Reissner-Nordstr\"om-Anti-de Sitter (RNAdS) black hole
\be
\label{RNAdS}
  a(r)\equiv 1 \ \ , \ \  f(r) = 1 + \frac{r^2}{L^2} - \frac{2M}{r^2} + 
\frac{\gamma Q^2}{r^4} \ \ , \ \ \phi(r) = \frac{Q}{r_h^2}\left(1 - \frac{r_h^2}{r^2}\right) \ .
\ee
The RNAdS black hole with horizon at $r=r_h$ has energy
\be
               E = \frac{V_3}{8 \pi G} 3 M \ \ , \ \ 
M = \frac{\gamma Q^2 + r_h^4 + r_h^6/ L^2}{2 r_h^2} 
\ee 
and these black holes exist for  $Q < Q_m$ and become  extremal for 
\be
\label{extremal}
                 Q_m = \frac{r_h^2}{L} \sqrt{\frac{L^2+2 r_h^2}{\gamma}}  \ .
\ee
Correspondingly, the black holes exist for 
\be
0 < r_h < r_{h,m}(Q)
\ee 
where the maximal value  $r_{h,m}$ is obtained by inverting (\ref{extremal}). 

Note that the RNAdS solution is a solution to the bosonic part of  $d=5$, ${\cal N}=8$ 
gauged supergravity with Lagrangian \cite{cvetic1,cvetic2}
\begin{equation}
\frac{{\cal L}}{\sqrt{-g_5}} = \frac{R}{16\pi G} + \frac{1}{L^2}V(\psi_1,\psi_2,\psi_3) -
\frac{1}{4} G_{ij} F_{\mu\nu}^i F^{\mu\nu,j} - \frac{1}{2} G_{ij} \partial_{\mu} 
\psi^i \partial^{\mu} \psi^j  + \frac{1}{48 \sqrt{-g_5}} \epsilon^{\mu\nu\rho\sigma\lambda} \epsilon_{ijk}
F_{\mu\nu}^i F_{\rho\sigma}^j A_{\lambda}^k \ ,  
\end{equation}
where 
\begin{equation}
 V(\psi_1,\psi_2,\psi_3) = 2 \sum_{i=1}^3 \psi_i^{-1}
\end{equation}
is the potential of the three real scalar fields $\psi_i$, $i=1,2,3$, which are
subject to the constraint $\Pi_{i=1}^3 \psi_i=1$. 
$F_{\mu\nu}^i$, $i=1,2,3$ correspond to the three Abelian field strength tensors, while
\begin{equation}
 G_{ij}=\frac{1}{2}{\rm diag}\left(\frac{1}{\psi_1^2}, \frac{1}{\psi_2^2}, \frac{1}{\psi_3^2}\right) \ .
\end{equation}
and $\sqrt{-g_5}$ is the determinant of the f\"unfbein.
This model has the following solution \cite{cvetic1,cvetic2}:
\begin{equation}
 ds^2 = -\left(H_1 H_2 H_3\right)^{-2/3} f dt^2 + 
\left(H_1 H_2 H_3\right)^{1/3} \left(f^{-1} d\tilde{r}^2 +
\tilde{r}^2 d\Omega^2_{3}\right)
\end{equation}
with
\begin{equation}
 f(r)=1-\frac{\mu}{\tilde{r}^2} + \frac{\tilde{r}^2}{L^2} H_1 H_2 H_3 \ \ , 
\ \ H_i = 1 + \frac{q_i}{\tilde{r}^2} \ \ , i=1,2,3 \ 
 \end{equation}
and scalar and gauge fields given by
\begin{equation}
 \psi^i=H_i^{-1} \left(H_1 H_2 H_3\right)^{1/3} \ \ , \ 
\ \phi_i\equiv A_t^i =\frac{Q_i}{\tilde{r}^2 + q_i}  \ ,
\end{equation}
where the $Q_i$ are the physical charges with $Q_i^2=q_i(1+q_i/r_h^2)
\left(r_h^2+\Pi_{j\neq i} (r_h^2 + q_j)\right)$ with $r_h$ the outer horizon radius.
The solution with $q_1=q_2=q_3\equiv q$ has trivial scalar fields $\psi_1=\psi_2=\psi_3\equiv 1$.
Using the new variable $r^2:=\tilde{r}^2 + q$ we find the form of the solution as given in
(\ref{RNAdS}). The aim of our paper is to consider scalar field condensation on such
a black hole, which in some sense would correspond to the onset of an instability related
to slightly non-equal charges $q_1$, $q_2$ and $q_3$ \cite{gubser_mitra}. Of course our field theoretical
model is not embedded into some supergravity model, but it can be seen as a toy model
of those.

\subsection{Static and charged black holes with scalar hair}
There exist no explicit solutions to the system of equations for
$\psi(r) \neq 0$. 
We have therefore used a numerical method to construct solutions.  
The field equations were solved by employing a collocation method for boundary-value ordinary
differential equations equipped with an adaptive mesh selection procedure \cite{colsys}.  
In the following we will use the rescalings and set $L=1$ and $\gamma=0.5$.

Note that this system has been studied for $G=0$ in \cite{gubser}, where
it was shown that a scalar field condenses in the background of the RNAdS solution. In \cite{dias2}
this system was studied, but for a massless scalar field. Here we are
looking at the full system of coupled differential equations with a tachyonic scalar field, which has before only been
studied for RNAdS solutions with flat horizon topology in the context of holographic
superconductors (see e.g. \cite{hhh,reviews} and references therein).

Our numerical results indicate that the black holes with scalar hair 
exist on a finite interval of the horizon radius $r_h$ for fixed values of $Q$ and $e$, i.e.
\be
\label{domain}
   r_{h,1} < r_h < r_{h,2}  \ \ {\rm with} \ \  r_{h,1} < r_{h,m}(Q) < r_{h,2}
\ee
where $r_{h,m}(Q)$ is the  maximal horizon radius of the RNAdS solution with the same value of
$Q$. $r_{h,1}$ and $r_{h,2}$ depend generally on $Q$ and $e^2$. 
In particular, both RNAdS solutions and the black hole solutions with scalar hair coexist
on some interval of $r_h$.
In the limit $r_h \to r_{h,2}$ the scalar field $\psi(r)$ and the metric field $a(r)$ 
approach uniformly  $\psi(r)\equiv 0$ and $a(r)\equiv 1$, 
while the metric function $f(r)$ and the
vector potential $\phi(r)$  approach the RNAdS solution with horizon $r_h=r_{h,2}$ and charge $Q$.

This is illustrated in Fig. \ref{fig1}, where the value of the ``condensate'' 
$\psi_+$, the value of the metric function $a(r)$ at the horizon, $a(r_h)$, as well as the
horizon radius $r_h$  
are given as functions
of the Hawking temperature $T_H$ for $e^2=1$ and $e^2=2$, respectively, 
for  several values of $Q$.
\begin{figure}[ht]
  \begin{center}
    \subfigure[$e=1$]{\label{fig1a}\includegraphics[scale=0.55]{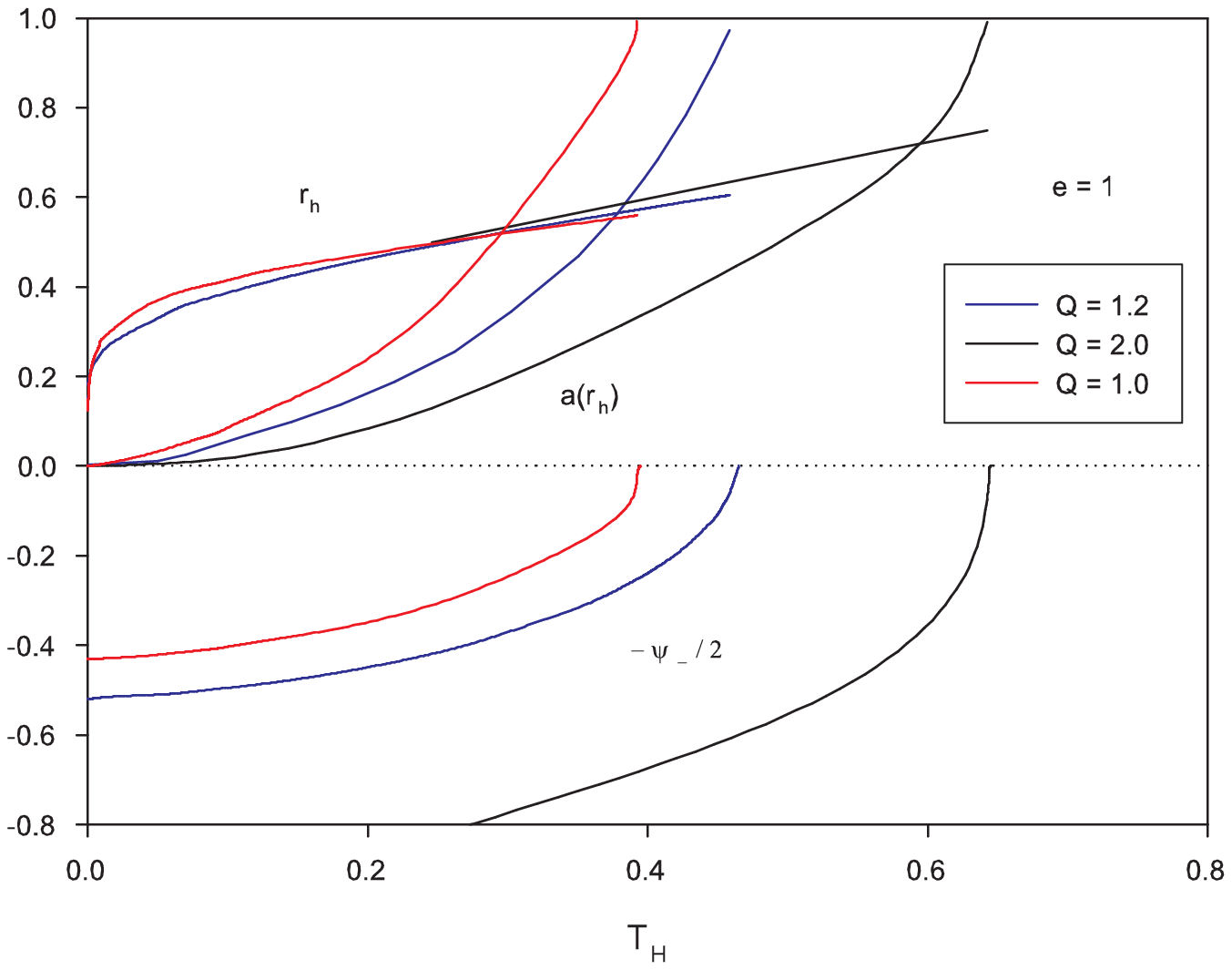}}
    \subfigure[$e^2=2$]{\label{fig1b}\includegraphics[scale=0.55]{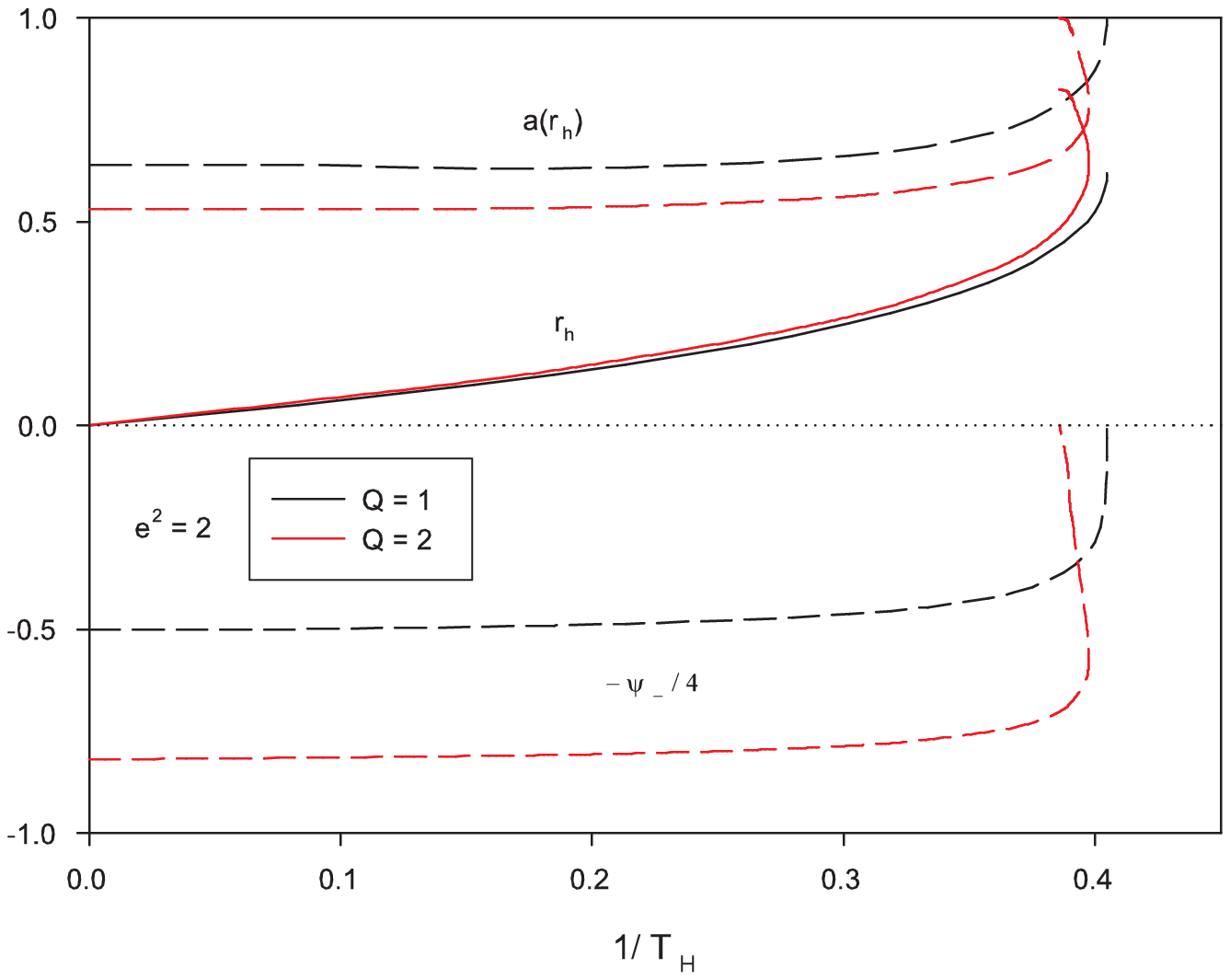}}
    \end{center}
   \caption{The ``condensate'' $\psi_+$, the value of the metric function $a(r)$ at the horizon, $a(r_h)$ and 
the horizon value $r_h$ are given as function of the Hawking temperature $T_H$ 
for $e^2=1$ and several values of $Q$ (left) and $e^2=2$ and two values of $Q$ (right), respectively.}
\label{fig1} 
 \end{figure}

We observe that the critical temperature as well as the horizon radius $r_h$ at which 
the scalar condensation
instability appears decreases with decreasing $Q$ for a fixed value of $e$.

Our results suggest further that the behaviour of the solutions in the
approach of the minimal value $r_{h,1}$ depends crucially on $e$. This has also
been observed for massless scalar fields \cite{dias2}.
For small values of $e$ (here $e=1$, see Fig.\ref{fig1a}) 
our numerical results suggest that black hole solutions exist for 
$r_h >0$ (i.e. $r_{h,1} = 0$), irrespectively of $Q$. 
Decreasing $r_h$ the value $a(r_h)$ approaches zero 
and our results suggest that it does so power-like. E.g. for $Q=e^2=1$
and small $r_h$ we find that our numerical data can be well fitted by
$a(r_h)\sim 200 r_h^8$. 
Accordingly, both the temperature $T_H$ and the entropy $S$ tend to zero.
The approach of this singular configuration renders the numerical integration difficult
and unreliable (since $a(r_h)$ becomes smaller than the tolerance we impose to control
the accuracy of the solutions). Howewer, all our results imply that the interpretation
$\lim_{r_h \to 0} a(r_h) = 0$ is correct.

For larger values of $e$, typically $e^2 > 1.5$ (here $e^2=2$, see Fig.\ref{fig1b}) however, the solutions exist all the way down to 
$r_h=0$. In this limit they merge with the soliton solutions discussed in \cite{basu,dias2}. 
This is clearly see in Fig.\ref{fig1b}, where it is shown that $a(r_h)$ and $\psi_+$ stay
finite in the limit $r_h\rightarrow 0$.

\begin{figure}
\centering
\epsfysize=8cm
\mbox{\epsffile{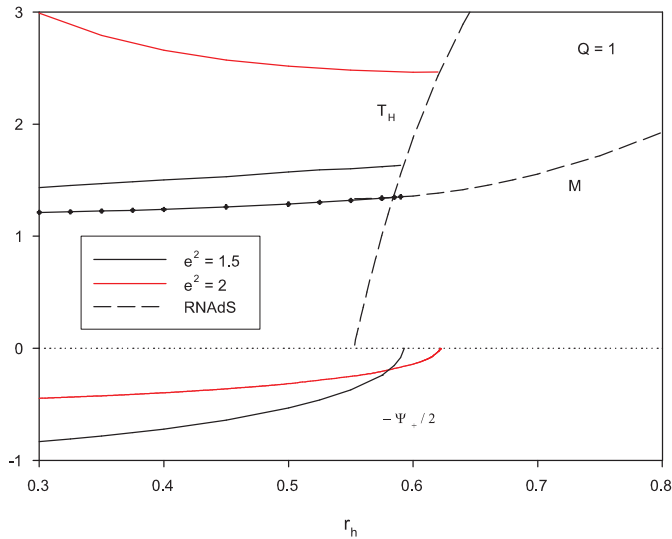}}
\caption{\label{F_TH_2}
The Hawking temperature $T_H$ and the ``condensate'' $\psi_+$ are shown in dependence on 
$r_h$ for non-rotating charged black holes with scalar hair. Here $Q=1$ and 
$e^2=1.5$ (black solid), $e^2=2$ (red solid), respectively. We also show the mass parameter $M$ 
for these solutions (solid line with dots).
For comparison, the Hawking temperature $T_H$ and the mass parameter $M$
of the RNAdS solution with $Q=1$ is also shown (dashed).}
\end{figure}
 
We have also studied the influence of the charge $e$ of the scalar field on the solutions
for fixed $Q$. Here we have chosen $Q=1$. 
The numerical results reveal that the pattern of solutions and the 
thermodynamical properties are strongly depending on
the parameter $e^2$. The results are given in Fig. \ref{F_TH_2}, where we
show the Hawking temperature $T_H$, the mass parameter $M$ as well as the ``condensate''
$\psi_+$ as a function of the horizon radius $r_h$. The mass of the solutions with
scalar hair deviates only little from that of the corresponding RNAdS solution with the same
charge. The curves for the temperature branch of from the curve of the temperature
of the RNAdS solution. We observe that the smaller $e^2$ the smaller $T_H$ (and $r_h$) at which
the scalar condensation instability appears.
The next question is then whether the solutions with or without scalar hair are thermodynamically
stable. For that we have plotted the entropy $S$ as a function of the temperature $T_H$ in 
Fig.\ref{TH_S}. We explore the solutions for constant charge $Q$. Accordingly, we work
in the canonical ensemble. As such the heat capacity 
$C = T_H \frac{\partial S}{\partial T_H}$ will tell us whether the solutions
are stable. For $C > 0$ (respectively $C <0$ the solutions are 
thermodynamically stable (respectively unstable).
If we had fixed the asymptotic value of $\phi$, i.e. $\mu$ we would
be working in the grandcanonical ensemble. In that case we would need to ensure that
both $C/T_H$ and 
$\left(\frac{\partial Q}{\partial \phi}\right)_{T_H}$ are positive in order to 
have thermodynamic stability \cite{monteiro}. 

First of all note that the RNAdS solutions are thermodynamically stable for large $Q$ (here $Q=1.0$), while
they become thermodynamically unstable for intermediate temperatures at small charge $Q$ (here $Q=0.1$).
The black hole solutions with scalar hair branch of from the curves for the RNAdS solutions with
the numbers next to the curves indicating the value of $e^2$. Note that the construction
of small black holes is numerically difficult -
 this is why the curves of the solutions with scalar hair stop. 
We observe that the smaller $e^2$ the smaller the temperature at which scalar condensation
appears. Moreover, for large $e^2$ the solutions with scalar hair have negative specific heat, i.e.
are unstable, while for small $e^2$ they have positive specific heat. The transition
value at which the specific heat changes from being negative to being positive
depends on the value of $Q$. The larger $Q$ the smaller we have to choose $e^2$ to find
thermodynamically stable solutions.

\begin{figure}
\centering
\epsfysize=8cm
\mbox{\epsffile{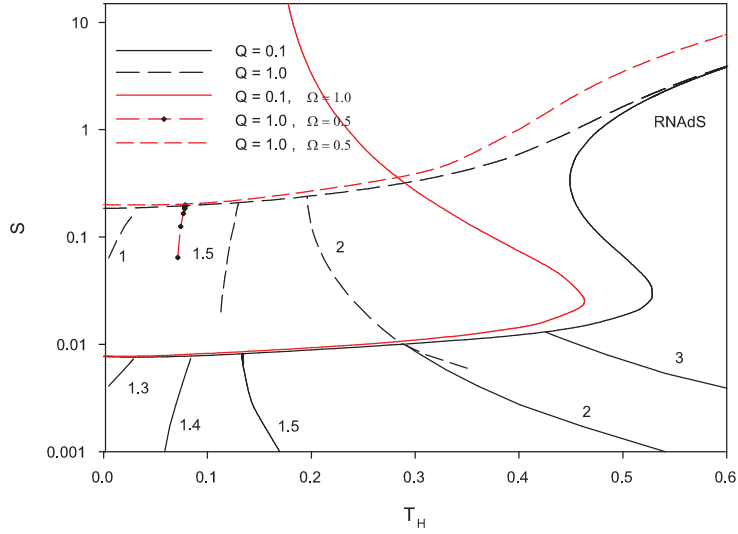}}
\caption{\label{TH_S}
The entropy $S$ as a function of the Hawking temperature $T_H$ for RNAdS solutions (black)
for $Q=0.1$ (solid) and $Q=1.0$ (dashed). The solid (resp. dashed) curves
branching of these main curves denote the entropy of the static charged black holes with
scalar hair. The numbers next to these curves denote the value of $e^2$. 
We also show the corresponding curves for the rotating charged black holes without scalar
hair (red) for $Q=0.1$, $\Omega=1.0$ (solid) and $Q=1.0$, $\Omega=0.5$ (dashed).
To compare with the static case, we also show the entropy of a rotating charged
black hole with scalar hair (dashed with dots) for $Q=1.0$, $\Omega=0.5$ and $e^2=1.5$, which
branches of the curve of the solutions without scalar hair. }
\end{figure}

We also plot the dependence of the entropy $S$ on the mass parameter $M$. This is given
in Fig.\ref{M_S_Q_1}. We observe that in the intervals in which the solutions with and
without scalar hair coexist, the solutions with scalar hair have higher entropy at a
given mass $M$ than the solutions without hair and are hence thermodynamically preferred.

\begin{figure}
\centering
\epsfysize=8cm
\mbox{\epsffile{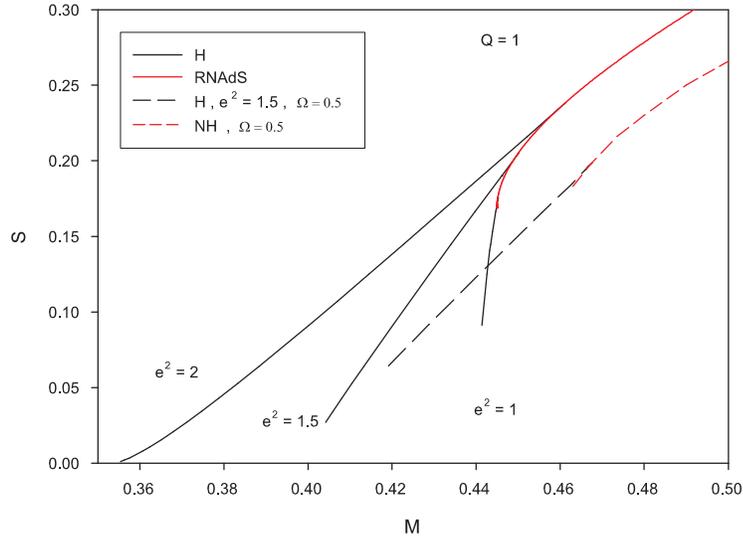}}
\caption{\label{M_S_Q_1}
Entropy $S$ as function of mass $M$ for RNAdS solutions (red solid line) and
rotating and charged black holes without scalar hair (NH, red dashed line) for $Q=1$.
We also show the entropy $S$ of the corresponding black hole solutions with scalar
hair for the non-rotating case for three values of $e^2$ (H, black solid lines)
as well as for the rotating case for $e^2=1.5$ and $\Omega=0.5$ (H, black dashed line).}
\end{figure}

\section{Rotating solitons and black holes}
In this case the Ricci scalar has the following form
\begin{equation}
 R=\frac{1}{g}\left(8 - \frac{2h}{g}\right) + \frac{f}{g b}b'g' + 
\frac{f h}{2 b } (w')^2 + \frac{f}{2g^2} (g')^2
       + \frac{f}{2b h} b' h' + \frac{f}{g h} g' h'
\end{equation}
and $\sqrt{-{\rm det(g_{\mu\nu})}}=g\sqrt{\frac{b h}{f}}$. 
The matter Lagrangian density reads
\begin{equation}
{\cal L}_{\rm matter} = {\cal L}_{s}  + {\cal L}_{g} 
\end{equation}
with 
\begin{equation}
{\cal L}_s = - m^2 \psi^2 - f (\psi')^2 + e^2 \psi^2 
\left(\frac{1}{b} \phi^2 + 2 \frac{w}{b} \phi A + \frac{h w^2-b}{bh} A^2\right)
\end{equation}
and
\begin{equation}
{\cal L}_g = \frac{f}{2b} (\phi'+w A')^2 - \frac{f}{2h} A'^2 -  \frac{2}{g^2} A^2 \ .
\end{equation}
The equations of motion can then be deduced from the variation of $S$ (see (\ref{action}))
with respect
to the metric and matter functions and we have seven 
coupled non-linear ordinary differential equations in this case.
The equation for $f(r)$ is first order, while the remaining equations are all
second order. These equations are quite involved - that is why we don't present them here.
Looking at the equation for $\psi$, we find that the effective mass of the scalar field now reads
\begin{equation}
 m_{\rm eff}^2=m^2 - \frac{e^2}{b} \left[
\phi^2 + A^2 \left(\omega^2-\frac{b}{h}\right) + 2 \phi A \omega\right] \ .
 \end{equation}
Choosing $m^2=-3/L^2 > m_{\rm BF}^2=-4/L^2$ the asymptotic AdS is again stable, but close
to the horizon, where $1/b$ can become very large, the effective mass can drop below the BF bound.
We would thus expect a scalar condensation instability close to the horizon of the rotating charged
black hole. This is what we will demonstrate in the following.

\subsection{Known solutions} 
\subsubsection{Rotating and uncharged black holes}
In the limit of vanishing gauge fields $\phi(r)\equiv 0$, $A(r)\equiv 0$ and vanishing
scalar field $\psi(r)\equiv 0$ the solution to the equations of motion are the AdS generalizations
\cite{HHT,GLP} of the Myers-Perry black holes
\cite{MP} with
\begin{equation}
 f(r)=1 + \frac{r^2}{L^2} - \frac{2M \Sigma}{r^2} + \frac{2Ma^2}{r^4} \ \ , \ \ 
h(r)=r^2\left(1+ \frac{2Ma^2}{r^4}\right) \
 \ , \ b(r)=\frac{r^2 f(r)}{h(r)} \ \ , \ \ \omega(r)=\frac{2Ma}{r^2 h(r)} 
\end{equation}
with $\Sigma=1-a^2/L^2$ and $M$ and $a$ are related to the energy and angular momentum of the
solutions as follows
\begin{equation}
 E= \frac{V_3}{8\pi G} M (4-\Sigma) \ \ , \ \ \bar{J}=\frac{V_3}{8\pi G} J \ \ {\rm with} \ \ J=2Ma \ .
\end{equation}

The instability of these solutions under condensation of a tachyonic scalar field
has been addressed in \cite{dias}. It was found that the black holes with scalar hair exist
only for very small temperatures with $T_H L \sim 10^{-3}$, i.e. condensation of the
scalar field is very hard if the rotating black hole is uncharged.
Here we want to address the condensation of a scalar field on a charged and rotating black
hole in (4+1)-dimensional AdS space-time. While exact solutions describing charged, rotating
black holes do exist in gauged supergravity models \cite{sugra_bh}, rotating black hole
solutions of pure Einstein-Maxwell theory are not known in closed form. These
have been constructed numerically in the asymptotically flat case \cite{kunz1}
as well as in AdS space-time for equal angular momenta \cite{kunz2} and with higher order
curvature corrections in \cite{brihaye1}. 
We will briefly discuss the solutions studied in \cite{kunz2} in the following before
presenting the solutions with scalar hair.

\subsubsection{Rotating and charged black holes without scalar hair}
Again, we have used a collocation method for boundary-value ordinary
differential equations equipped with an adaptive mesh selection procedure \cite{colsys} to solve
the system of coupled, nonlinear ordinary differential equations.

In order to understand the pattern of rotating charged solutions that possess scalar hair
let us first reinvestigate the domain of existence of the rotating charged black holes
without scalar hair in the parameter space $(\Omega,r_h)$ with $Q$ fixed.
As mentioned above the system has been studied previously \cite{kunz2,brihaye1} and it
was found that for fixed $Q$ black holes exist
in a domain   
\begin{equation}
r_h  \geq  r_{h,m}(\Omega,Q)  \ \ {\rm or} \ \  0  \leq \Omega \leq \Omega_{max}(r_h,Q) \ .
\end{equation}
The limiting solution at $r_h=r_{h,m}(\Omega,Q)$ is an extremal black hole solution.
For non-rotating solutions the value of $r_{h,m}(0,Q)$ is non-vanishing and can be determined 
analytically from (\ref{extremal}). For rotating solutions the value of $\Omega_{max}(r_h,Q)$
has to be determined numerically.
Our numerical results indicate that the  value $\Omega_{max}(r_h,Q)$   
is a monotonically increasing function of $r_h$ and that in the limit $\Omega \to 0$,
the horizon radius of the extremal RNAdS solution (with the corresponding $Q$) is approached.
The numerical determination of  $\Omega_{max}(r_h,Q)$ is, however, difficult because of the
fact that the limiting solution is an extremal black hole.
The dependence of the Hawking temperature $T_H$, the mass parameter $M$ and the angular
momentum $J$ (in units of $V_3/(8\pi G)$) on $\Omega$ is given in Fig.\ref{physical} for $Q=1$ and 
$r_h=0.56$ (blue curve) and $r_h=0.575$ (red curve), respectively.
For $Q=1$, the non-rotating charged solution exists down to $r_h \approx 0.55$.
The solution with $\Omega=0$ and $r_h=0.56$ is therefore 
very close to the
extremal limit and the black hole can be made spinning up to $\Omega_{max} \approx 0.35$.
The mass parameter $M$ as well as the angular momentum increase with increasing $\Omega$, while
the temperature $T_H$ decreases. We find that the larger the black hole horizon, the 
bigger we can choose $\Omega$: for $r_h=0.575$ we find $\Omega_{max}\approx 0.63$ (see Fig.\ref{physical}),
while for $r_h=2.0$ we have $\Omega_{max} \approx 1.1$.

\subsection{Rotating and charged solutions with scalar hair}
Again, the solutions have to be constructed numerically and we employed the collocation
method mentioned earlier in this paper \cite{colsys}.  
\subsubsection{Rotating and charged solitons with scalar hair}
In \cite{basu,dias2} static soliton solutions with scalar hair  have been
constructed in the case of a massless scalar field. It was found that for small
gauge coupling the solutions exist up to some critical value of the charge and
that at this critical value of the charge the solution becomes singular. On the other
hand for sufficiently large gauge coupling the solutions exist up to infinite charge.
Moreover, it was noted that for a given value of small enough gauge coupling the charge does
not uniquely specify the solution.

The question is then whether within our Ansatz, we can also construct rotating soliton
solutions with scalar hair. 
For that let us consider the boundary conditions at $r=0$. We find that we need to impose
\begin{equation}
f(0) = 1 \ , \ h(0) = 0 \ , \ A(0) = 0 \ , \ \omega(0)=\Omega \ \ . 
\end{equation}
Moreover, the derivatives of all seven functions $f$, $b$, $h$, $\omega$, $\phi$, $\psi$, $A$
need to vanish at $r=0$. 
In addition, the boundary conditions (\ref{bcads1}), (\ref{bcads2}) on the AdS boundary 
need to be fulfilled. Imposing these conditions, we find that the only possible solution
is given by
\begin{equation}
    h(r) = r^2 \ \ , \ \  \omega(r) = \omega(0)= \Omega \ \  , \ \ A(r) = 0 \ 
\end{equation}
and the other functions approach those of the static hairy soliton solution discussed in \cite{basu,dias2}.
This is nothing else but a globally rotated hairy soliton solution with vanishing
angular momentum $\bar{J}$. Hence, our results
strongly suggest that within our Ansatz there are no non-trivial rotating and charged hairy solitons.
In Fig.\ref{reg_rot} we show a soliton solution together with a corresponding
black hole solution for $e^2=2$, $Q=1$ and $\Omega=0.3$. 

\begin{figure}
\centering
\epsfysize=8cm
\mbox{\epsffile{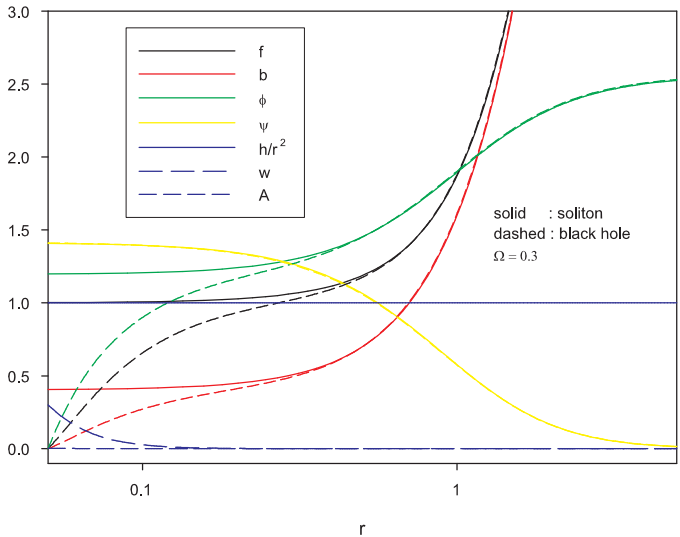}}
\caption{\label{reg_rot}
We show the metric and matter functions for a hairy soliton solution with
$e^2=2$, $Q=1$ and $\Omega=0.3$ (solid). For comparison we also show a corresponding
black hole solution for $r_h=0.05$ close to the soliton solution (dashed).}
\end{figure}

Since in comparison to \cite{dias2} our scalar field is tachyonic, we also present
some results on the soliton solutions here and remark that these are valid for
any $\omega(r)=\Omega$ including, of course, the case $\Omega=0$. 
Our results are presented in Fig.\ref{reg_e2}, where we give the value of the metric
function $a(r)=\sqrt{b(r)/f(r)}$ (see also the static case and the corresponding
equation (\ref{eq2})) and the gauge field function $\phi(r)$ at $r=0$, the value of
the ``condensate'' $\psi_+$, the value of $\mu$, i.e. $\phi(\infty)$ and the
mass parameter $M$ as function of $e^2$ for two different values of $Q$.
We observe that the solutions exist down to some
minimal value of $e^2=e^2_{\rm min}(Q)$ and that at this value of $e^2$ a second branch
of soliton solutions emerges from the first branch. This second branch exists
up to some critical 
$e^2=e^2_{\rm c}(Q) > e^2_{\rm min}(Q)$ at which the soliton solutions become
singular with $a(0)\rightarrow 0$. This approach to a singular solution was also observed for
massless scalar fields in \cite{dias2}.

\begin{figure}[ht]
  \begin{center}
    \subfigure[]{\label{reg_e2_b}\includegraphics[scale=0.55]{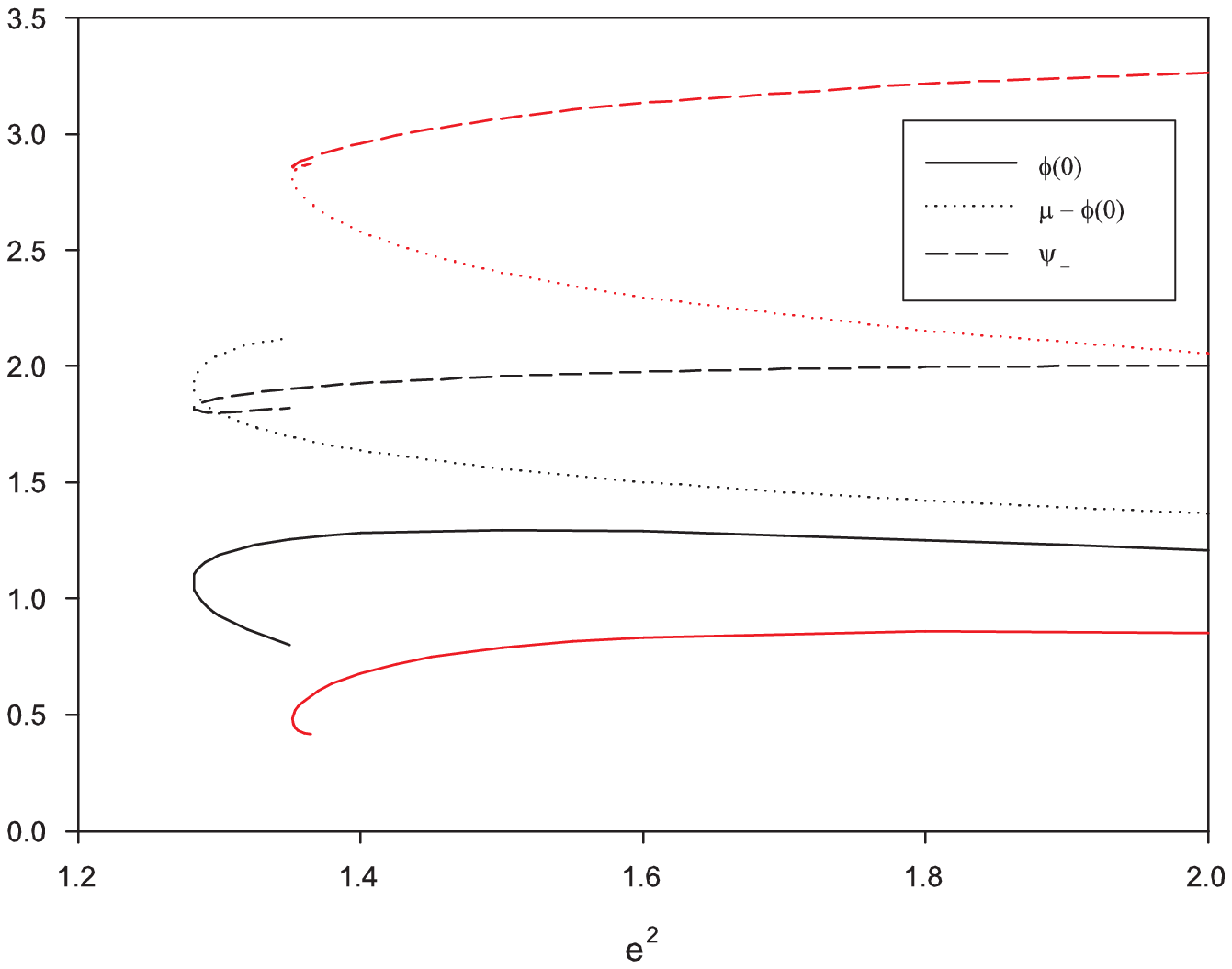}}
    \subfigure[]{\label{reg_e2_a}\includegraphics[scale=0.55]{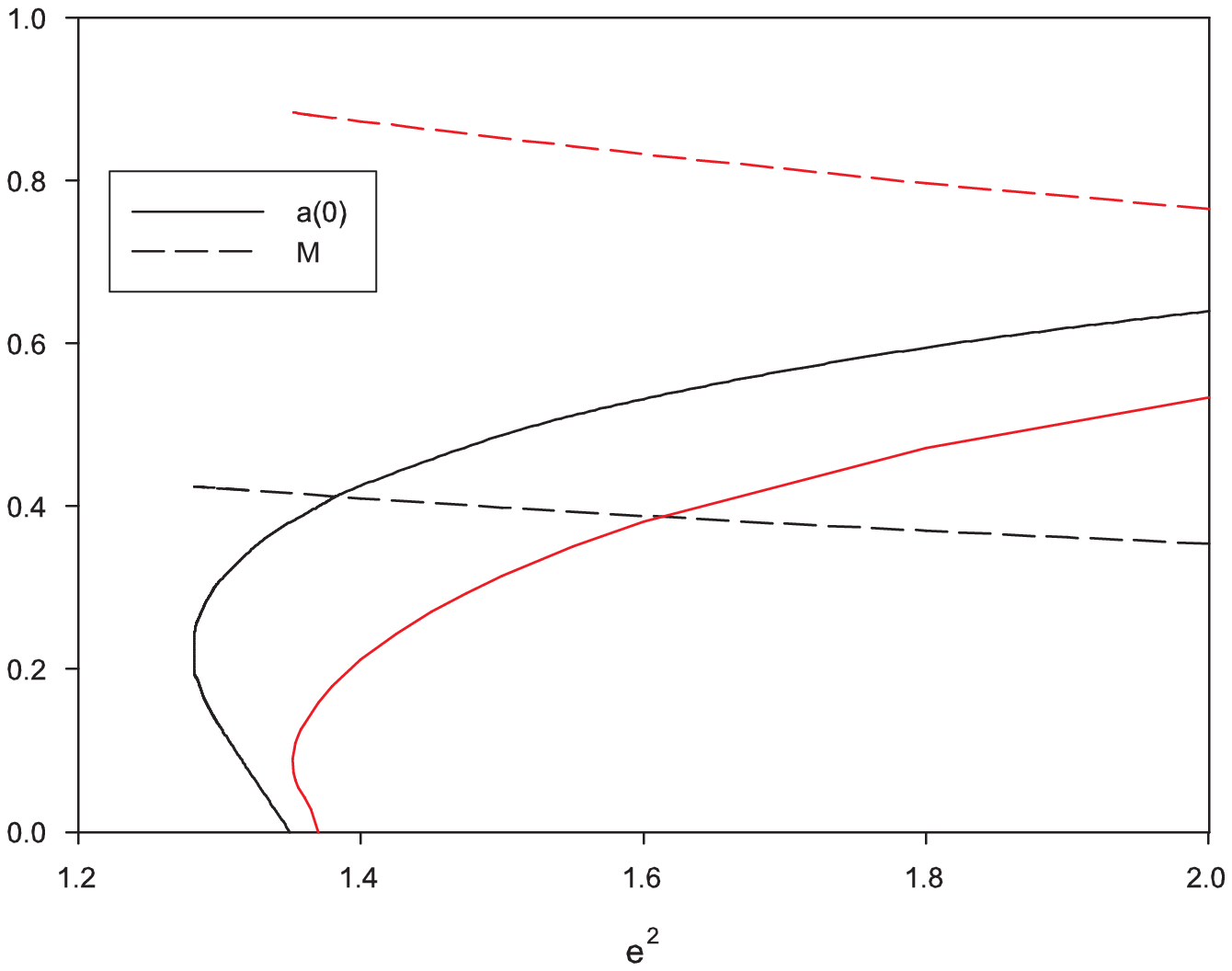}}
    \end{center}
   \caption{The value of the potential
$\phi(r)$ at $r=0$ (solid), the value of the chemical potential $\mu$ (dotted)
and the value of the ``condensate'' $\psi_+$ (dashed) (a) as well as the
metric function $a(r)$ at $r=0$ (solid) and the mass $M$ (dashed) (b) are given as function
of $e^2$ for soliton solutions with $Q=1$ (black) and $Q=2$ (red), respectively. }
\label{reg_e2} 
 \end{figure}
Hence, on the interval $[e^2_{\rm min}:e^2_{\rm c}]$ two soliton solutions exist for the
same values of $Q$ and $e^2$. However, the solutions can be uniquely labeled by the value of
one of the matter or metric functions at $r=0$ or on the AdS boundary. From Fig.\ref{reg_e2_b} we
see that $\phi(0)$ as well as $\psi_+$ are both smaller on the second branch of solutions
for a fixed value of $e^2$ and $Q$. On the other hand, the chemical potential $\mu=\phi(\infty)$ is
larger on the second branch of solutions for fixed $e^2$ and $Q$. Moreover, from
Fig.\ref{reg_e2_a} we find that $a(0)$ is smaller on the second branch, while
the mass parameter $M$ is larger. This is not clearly seen on the figure because
the masses are very close to each other (within 0.1$\%$ of the absolute value), but our
numerical accuracy allows us to make the mentioned observation.

We find that both $e^2_{\rm min}$ as well as $e^2_{\rm c}$ depend on $Q$ and 
increase with increasing $Q$: $e^2_{\rm min}(Q=1)\approx 1.28$, $e^2_{\rm min}(Q=2)\approx
1.35$ and  $e^2_{\rm c}(Q=1)\approx 1.35$, $e^2_{\rm c}(Q=2)\approx 1.37$.

\subsubsection{Rotating and charged black holes with scalar hair}
As mentioned above we would expect rotating charged black holes to exhibit a 
scalar condensation instability in some domain of the parameter space. The corresponding
solutions would be rotating charged black holes with scalar hair. In the following
we will describe our numerical results for these solutions.

In Fig.\ref{physical} we present the mass parameter $M$, the temperature $T_H$, the angular momentum
$J$ and the ``condensate'' $\psi_+$ of a rotating charged black hole with scalar hair
for $Q=1$, $r_h=0.575$ and $e^2=1.5$ in dependence on $\Omega$. 
Note that for this choice of parameters the solutions with and without scalar hair coexist.
Comparing these two solutions we observe that the values of $M$ and $J$ are very close to
each other (within $1\%$ of the absolute value). Our numerical results indicate
within our numerical accuracy that the mass parameter $M$ of the solution with scalar hair 
is slightly lower
for all values of $\Omega$. Moreover, the solutions with scalar hair exist for larger values
of $\Omega$, $J$ and $M$. In contrast to this, the curves showing the dependence of the temperature
on $\Omega$ can be easily distinguished. The solutions with scalar hair have higher temperature
for a given value of $\Omega$. Moreover at $\Omega_{max}$ the value of the condensate
$\psi_+$ stays finite. 
The profile of a fast rotating (i.e. $\Omega$ close to $\Omega_{max}$) charged black hole with scalar hair
is  presented 
in Fig. \ref{fast_rotating} for $e^2=1.5$, $r_h=0.5$ and $\Omega = 0.75$.
The metric function $f(r)$ of the corresponding non-rotating solution is given by the dashed line
on Fig. \ref{fast_rotating_1}. We observe that 
$b/f$ is finite in the zero temperature limit, while the profile of $f$ 
suggests that $f'\vert_{r_h}=0$. 
The limiting solution, however, is not a regular extremal
hairy black hole solution. 
This was already observed before for rotating uncharged solutions 
\cite{dias} as well as for non-rotating charged solutions \cite{horowitz1}.
As found in \cite{fiol1} the scalar field should vanish on the horizon if the limiting
solution were a regular extremal black hole with scalar hair. However, we find
that the value of the scalar field on the horizon $\psi(r_h)$ increases
for decreasing temperature. Hence, the limiting solution cannot be a regular extremal
black hole with scalar hair. 

\begin{figure}
\centering
\epsfysize=8cm
\mbox{\epsffile{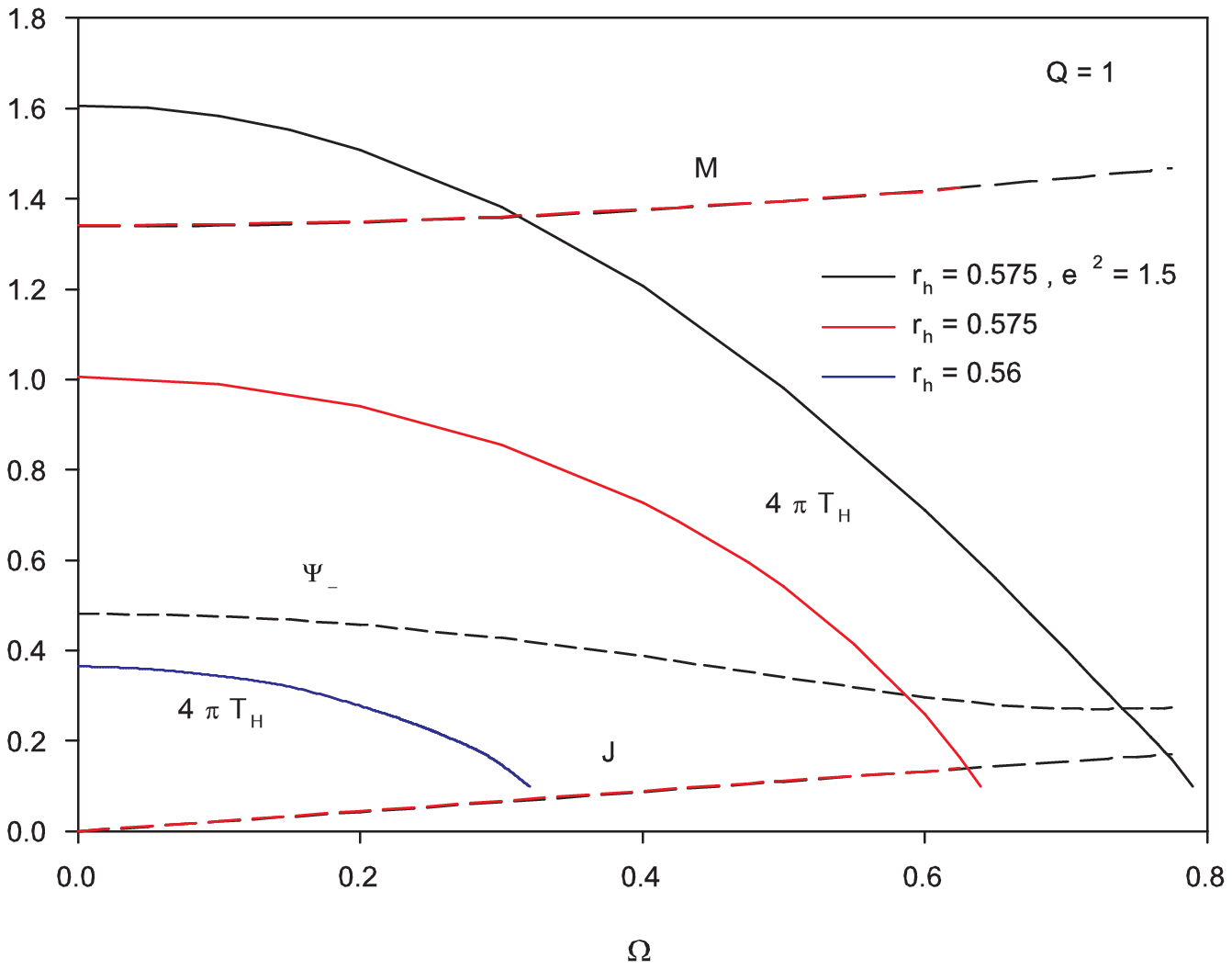}}
\caption{\label{physical}
We show the mass parameter $M$ (long-dashed), Hawking temperature $T_H$ (solid)  and 
angular momentum $J$ (short-dashed) 
of rotating charged black holes without scalar hair and $Q=1$ as function of $\Omega$.
Here we choose $r_h=0.56$ (blue) and $r_h=0.575$ (red), respectively. We also give the
same data for rotating charged black holes with scalar hair for
$r_h=0.575$, $e^2=1.5$ and $Q=1$ (black). Additionally we present the value of the ``condensate''
$\psi_+$ for these latter solutions.}
\end{figure}

\begin{figure}[ht]
  \begin{center}
    \subfigure[Metric functions]{\label{fast_rotating_1}\includegraphics[scale=0.55]{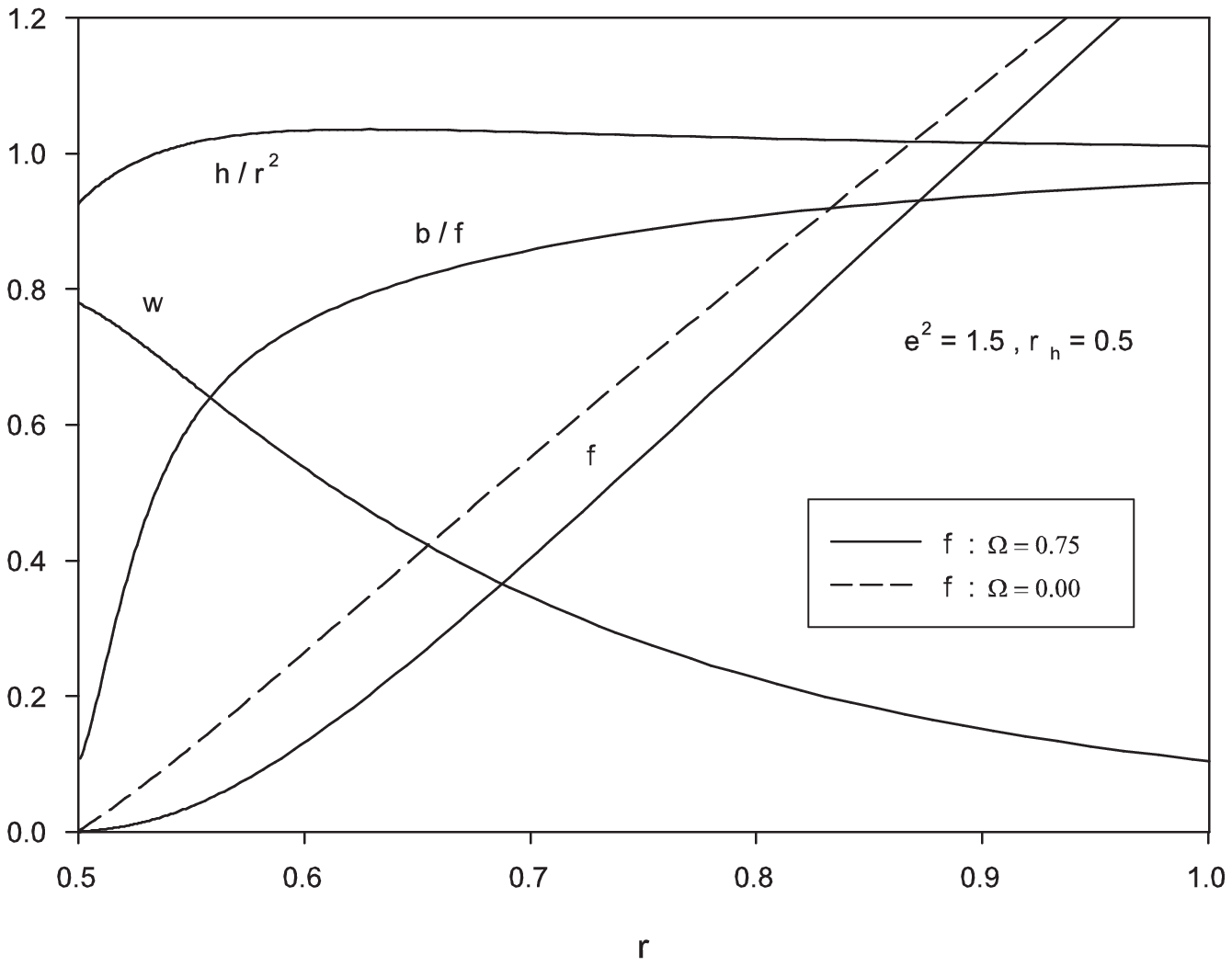}}
    \subfigure[Matter functions]{\label{fast_rotating_2}\includegraphics[scale=0.55]{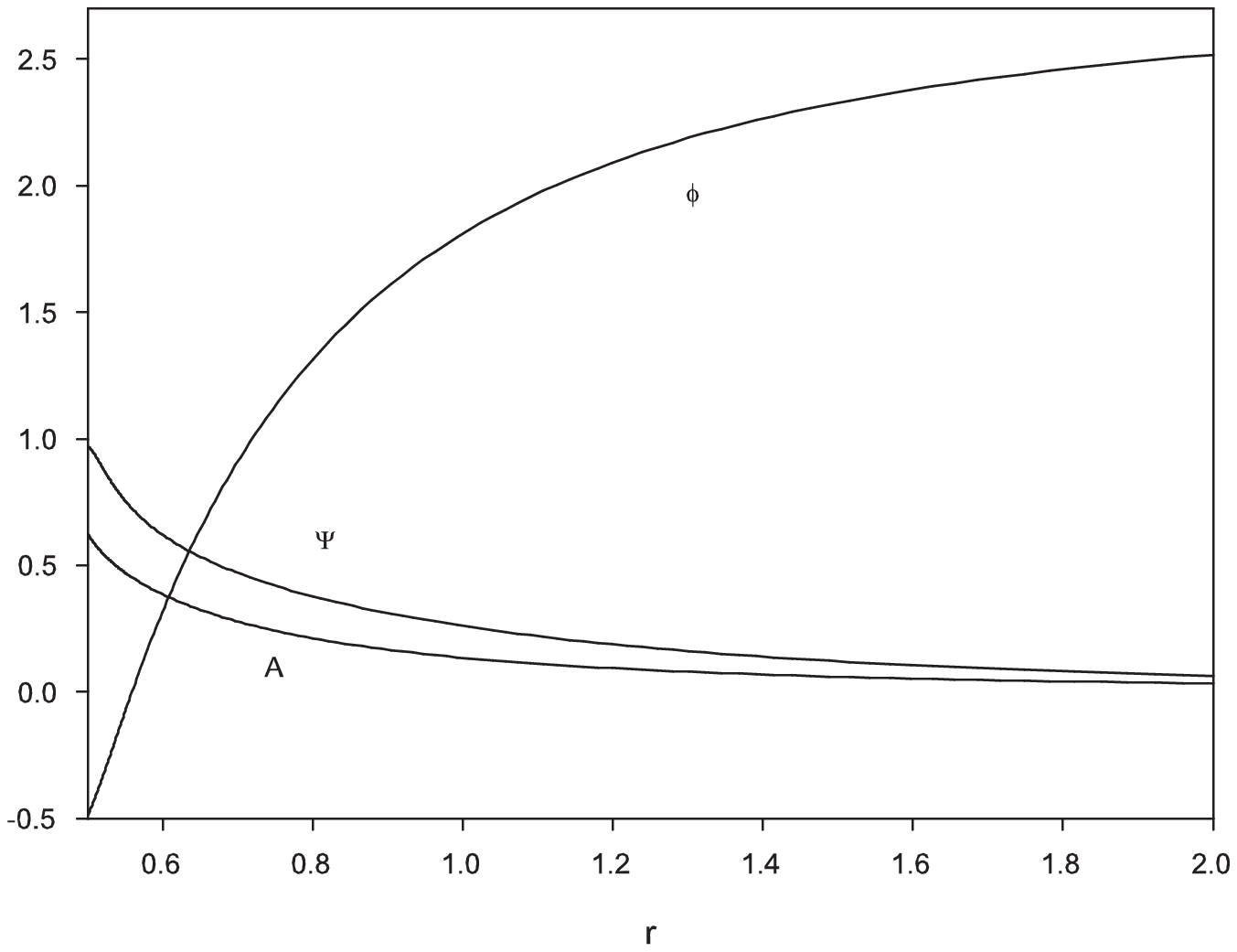}}
    \end{center}
   \caption{A fast rotating charged black hole with scalar hair 
for $e^2=1.5$, $r_h=0.5$ and $\Omega = 0.75$ close to $\Omega_{max}$. We show the
profiles of the metric functions $f$, $g$, $b$ and $\omega$ (left) and
of the matter functions $\phi$, $A$ and $\psi$ (right). The dashed line in the
left figure represents the metric function $f$ of the corresponding non-rotating solution.}
\label{fast_rotating} 
 \end{figure}

We have also studied the dependence of the critical temperature, i.e. the temperature at
which the black hole becomes unstable to scalar condensation in dependence on $\Omega$.
We find that for fixed $Q$
the critical temperature decreases with $\Omega$. This can e.g. be seen when looking at Fig.\ref{TH_S}
and comparing the curves that branch of from the RNAdS for $e^2=1.5$. 
Looking at Fig.\ref{M_S_Q_1} it is also obvious that the rotating charged black hole with
scalar hair is thermodynamically preferred over the solution without scalar hair since
in the mass interval in which they coexist the entropy $S$ of the solution with scalar
hair is larger.

Depending on this choice
we find that the solutions either tend to a singular solution that is not a regular extremal
black hole or to a soliton solution with $r_{h,1}=0$. 

Considering the black holes for varying horizon radius, we find that the black holes with scalar hair
again exist on a finite interval of the horizon radius $r_h$ with $r_{h,1} < r_h < r_{h,2}$.
In the limit $r_h \rightarrow r_{h,2}$ the solutions tend to the solutions without scalar hair,
i.e. they have $\psi(r)\equiv 0$. In the limit $r_h\rightarrow r_{h,1}$ our numerical results
indicate that the behaviour again depends on the values of $e$ and $Q$, but does
not to seem to depend on the value of the angular momentum of the solutions.
If for a choice of $e$ and $Q$ a soliton solution exists (see Fig.\ref{reg_e2}) the rotating
and charged black hole with $\omega(r_h)=\Omega$ will tend to
a soliton solution with $\omega(r)\equiv \Omega$. For a given $Q$ this is the
case for $e^2 > e^2_{\rm min}$. For a given $Q$ and $e^2 < e^2_{\rm min}$ no soliton
solution exists and the rotating black hole becomes a singular rotating solution
with vanising temperature at $r_h \rightarrow r_{h,1} > 0$. Our numerical
results indicate that for fixed values of $e$ and $Q$  the interval $[r_{h,1}:r_{h,2}]$ 
shrinks with increasing $\Omega$.

\section{Conclusions}
In this paper we have studied the condensation of a tachyonic scalar field on charged black hole
solutions in (4+1)-dimensional AdS space-time. 
We have investigated both the non-rotating as
well as the rotating case and have constructed black holes with scalar hair 
numerically for both cases. We find that the solutions exist on a finite
interval of the horizon radius $r_h$ and that the solutions tend to extremal
black hole solutions without scalar hair when approaching the maximal
horizon radius. On the other hand when approaching the minimal horizon
radius the solutions exist either all the way down to $r_h=0$ and are hence
soliton solutions with scalar hair or they tend to a singular solution
at a finite horizon radius. This latter solution is not a regular extremal
black hole and hence neither static nor rotating extremal black holes with scalar
hair exist. For the rotating case, one could think that the limiting
soliton solution is a rotating soliton with scalar hair. However, investigating
the boundary conditions we conclude that the limiting soliton solution
has vanishing angular momentum. This is confirmed by our numerical results.

Condensation of a charged scalar field on planar Gauss-Bonnet black holes
has been investigated in the context of
holographic superconductors \cite{gregory1,hartmann_brihaye4,gregory2}.
In \cite{hartmann_brihaye3} the condensation of an uncharged scalar field on
a hyperbolic Gauss-Bonnet black hole was studied. Rotating black holes in
Gauss-Bonnet gravity without scalar hair have recently been constructed for the
uncharged case \cite{brihaye_radu} as well as for the charged case \cite{brihaye1}.
It would be interesting to see whether these black holes exhibit a scalar
condensation instability and how the inclusion of the Gauss-Bonnet term alters the results
presented here. \\
\\

{\bf Acknowledgments} YB thanks the Belgian FNRS for financial support.

\end{document}